# New Directions in International Masterclasses

K. Cecire, on behalf of the IPPOG and QuarkNet Collaborations
*University of Notre Dame, Notre Dame, IN 46556 USA*

R. Dower
*Roxbury Latin School, West Roxbury, MA 02132*

International Masterclasses (IMC) have developed since their introduction in 2005. Masterclasses for International Day of Women and Girls in Science (IDWGS) and World Wide Data Day (W2D2) are innovations that began two years ago and are going well. With IDWGS masterclasses, a new pathway has been opened for high school girls to be encouraged in physics. W2D2 establishes new ways to bring masterclass activities directly to high school classrooms. New masterclass measurements beyond those for the LHC have been developed and tested, notably the MINERvA neutrino masterclass, which is the first IMC offering in neutrino physics and the first based on a Fermilab experiment. In the MINERvA measurement, students are able to study interactions of a neutrino beam with carbon nuclei, using conservation of momentum to draw conclusions. Other masterclass measurements related to Belle II and medical imaging are also in the testing stage. More neutrino masterclasses are in development as well, especially for MicroBooNE. A longer-term goal is the creation of a DUNE masterclass measurement as that facility reaches the data-taking stage.

## 1. INTERNATIONAL MASTERCLASSES

### 1.1. The Masterclass Concept

International Masterclasses enable high school students to work with physicist mentors becoming "particle physicists for a day." Using short QuarkNet activities, teachers can prepare their students and bring them to a university or a research institution nearby. Students learn more about particle physics from a mentor before making a measurement based on authentic data from a contemporary experiment. At the end of the day, masterclass participants attend a videoconference with physicists at CERN or Fermilab, teachers, and students from other locations, often in other countries, who have made the same measurement. For neutrino masterclasses, scientists at Fermilab moderate the videoconferences. International Masterclasses are held in and near March each year.

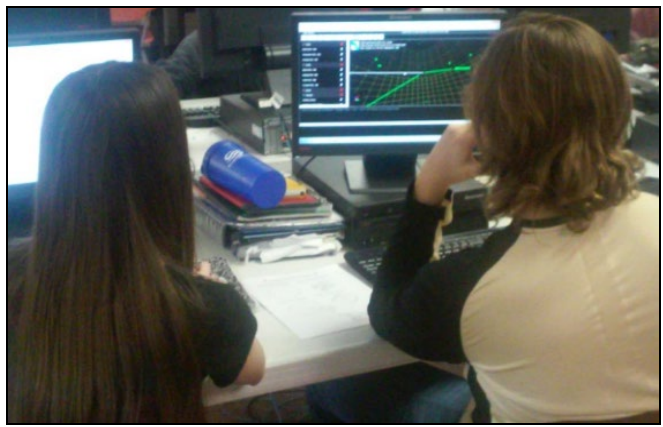

Figure 1: Students examine data from a CMS event display.

The International Masterclass day is structured around the following elements:
- Introductory presentation
- Tour of laboratory facilities
- Analyzing authentic HEP data
- Videoconference

## 1.2. International Masterclasses in Practice

International Masterclasses began in 2005 in Europe, a project of the European Particle Physics Outreach Group (EPPOG) with a small number of institutions. By 2019, now under the auspices of the International Particle Physics Outreach Group (IPPOG, successor to EPPOG), it is worldwide in scope with over 50 countries represented. In 2019, over 10,000 high school students participated in 332 masterclasses at 239 institutions. They made measurements of authentic data from four Large Hadron Collider (LHC) experiments at CERN and one experiment each at Fermilab and KEK. Moderation of videoconferences was shared by CERN, Fermilab, TRIUMF, and KEK, with the largest number of these at CERN (69) and Fermilab (27). [1]

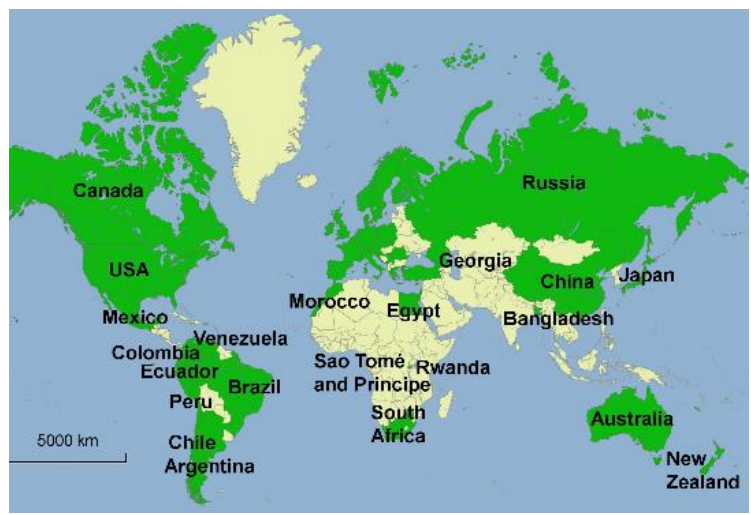

Figure 2: Countries in green participate in International Masterclasses.

## 2. SPECIAL MASTERCLASS PROJECTS

International Masterclasses have begun a new type of growth. Expansion continues in the number of masterclasses and in the geographical range of masterclass institutions. In addition, there are new opportunities for students and teachers to engage in masterclasses and new experimental particle physics measurements that go beyond the mainstay masterclass measurements from the LHC. The new opportunities are seen in the two special projects of International Masterclasses Central Coordination: World Wide Data Day and masterclasses for the International Day of Women and Girls in Science.

## 2.1. World Wide Data Day

World Wide Data Day (W2D2) is a single day of videoconferences which conclude student and teacher measurements that could have been done that day or any day in the weeks leading up to it. The purpose is to have a compact masterclass-style experience that can be done completely from school, without the need to visit a masterclass institution or even involve a physicist locally. Teachers help students to analyze CMS or ATLAS events and look specifically for dimuon events. These events are easy to spot even without making any cuts or other adjustments to the event display. Students then measure the direction angles θ and ϕ of each muon by setting the display to the z-y and x-y views of the event, respectively, and using a protractor. Students bin the results into 20º intervals on a tally sheet; the class uses the tally sheets to make class histograms of θ and ϕ and to enter their results into a Google spreadsheet which displays histograms showing the combined results worldwide. The real physics discussion takes place in the W2D2 videoconference. [2]

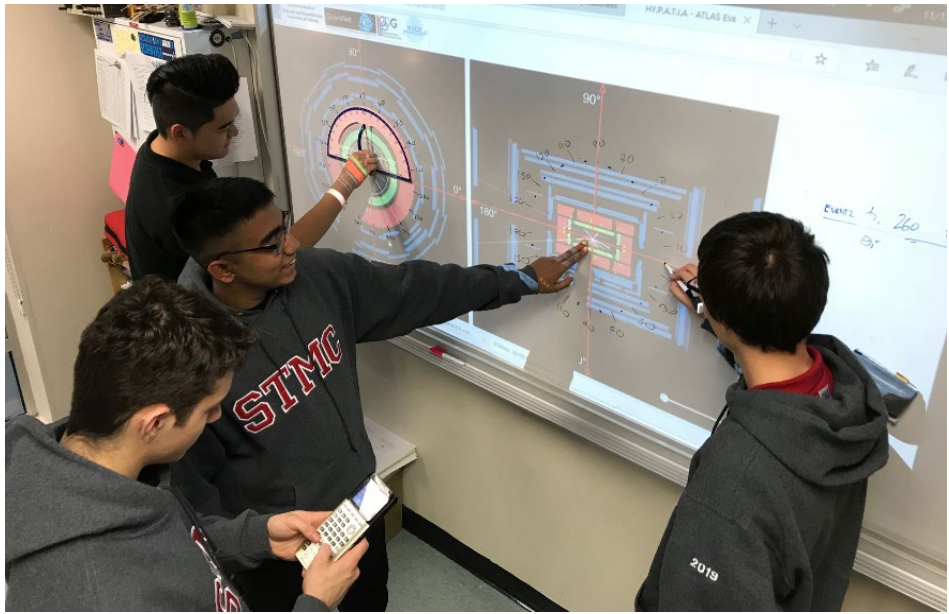

Figure 3: Students in Burnaby, British Columbia measure muon angles in ATLAS.

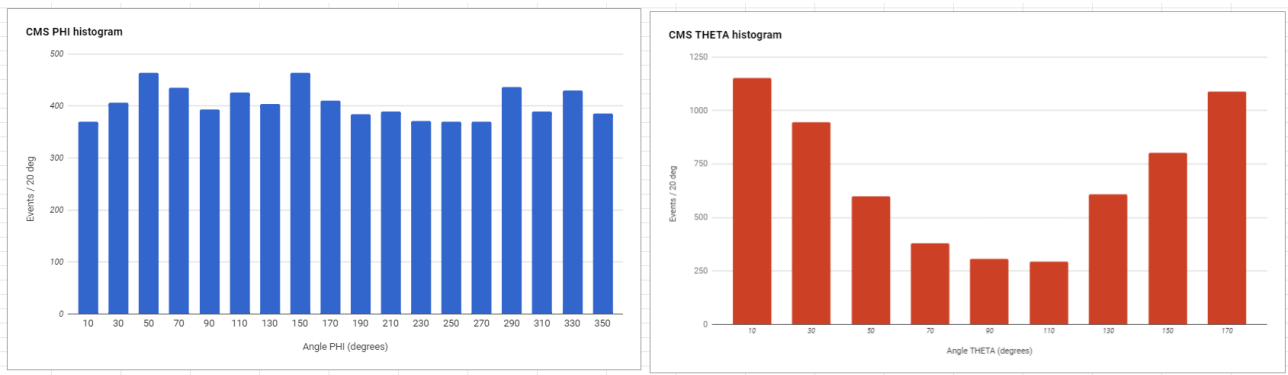

Figure 4: Global CMS results for muon angles ϕ and θ in World Wide Data Day 2018.

W2D2 was piloted in December 2016; it was held again in November 2017 and 2018 and will be held this year on October 16. It has grown each year in terms of numbers of students, schools, and countries. In 2018, over 1,000 students at 66 schools in 19 countries participated. In addition, because the W2D2 measurements are simple and compact, they have been used in workshops for large numbers of students in Mexico, Namibia, and Chile.

## 2.2. International Day of Women and Girls in Science

Since 2016, February 11 each year has been the United Nations International Day of Women and Girls in Science (IDWGS). [3] On and around that day each year, special masterclasses for women and girls have been held, united by CERN-moderated videoconferences which focus not only on the physics but also on issues and opportunities for women in science.

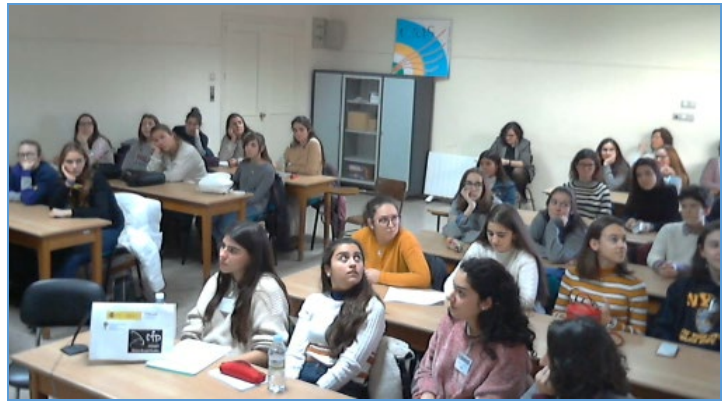

Figure 5: Young women in an IDWGS masterclass in 2019.

The IMC effort for IDWGS emphasizes masterclasses in a friendly physics environment for young women and women as role models for the students. Most of the participants are high school girls, and women are well represented among the masterclass tutors and organizers. The moderators at CERN are women. In 2018, there were 12 IDWGS masterclasses in five European countries plus Brazil.

## 3. NEW MASTERCLASS MEASUREMENTS

International Masterclasses have been focused, since 2005, on the LHC, first with Large Electron-Positron Collider masterclasses in anticipation of the LHC and then, starting in 2010, with LHC data from ALICE, ATLAS, CMS, and LHCb. Since then, some non-LHC, non-CERN groups began to develop masterclasses, notably IceCube and Pierre Auger. In 2018, development of "other-brand" measurements for International Masterclasses began in earnest. Belle II (KEK) physicists began work on the Belle II masterclass. GSI Darmstadt and the Heidelberg Ion Therapy Center began to pioneer a particle therapy masterclass. QuarkNet and Fermilab collaborated to develop and roll out accelerator–based neutrino masterclasses with the ultimate goal, still years off, of producing a masterclass for the Deep Underground Neutrino Experiment (DUNE).

## 3.1. Upgrade of the CMS Masterclass

The CMS masterclass has come in two varieties since 2011. The first is the original CMS J/Ψ measurement used in International Masterclasses in 2010. As much more dilepton data became available from CMS, the QuarkNet developers went to a W and Z masterclass, which included not only these boson but also relatively low-mass mesons that decay into lepton pairs, most notably J/Ψ and Y. After the Higgs discovery, a small number of four-lepton and diphoton Higgs candidate events became available and were incorporated to make the current CMS WZH measurement.

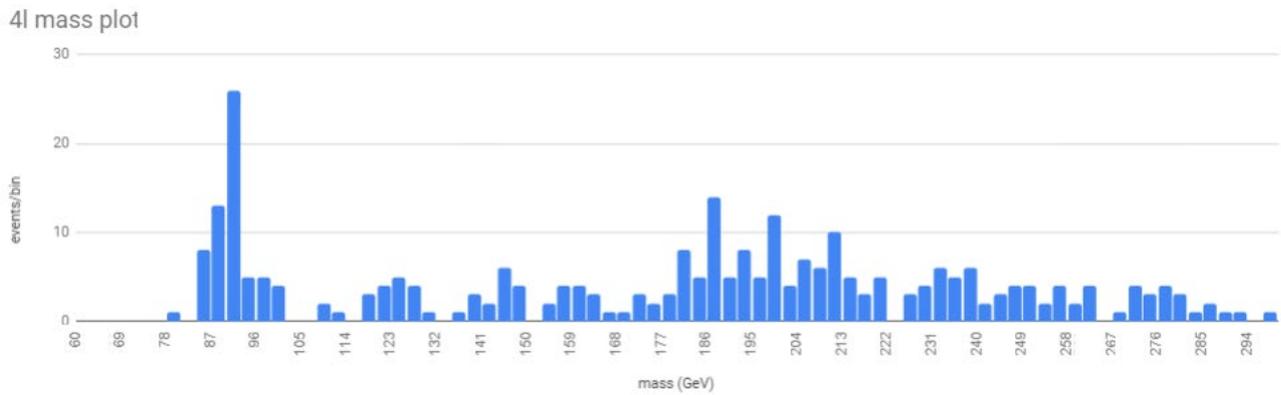

Figure 6: Test version of CMS masterclass four-lepton mass plot.

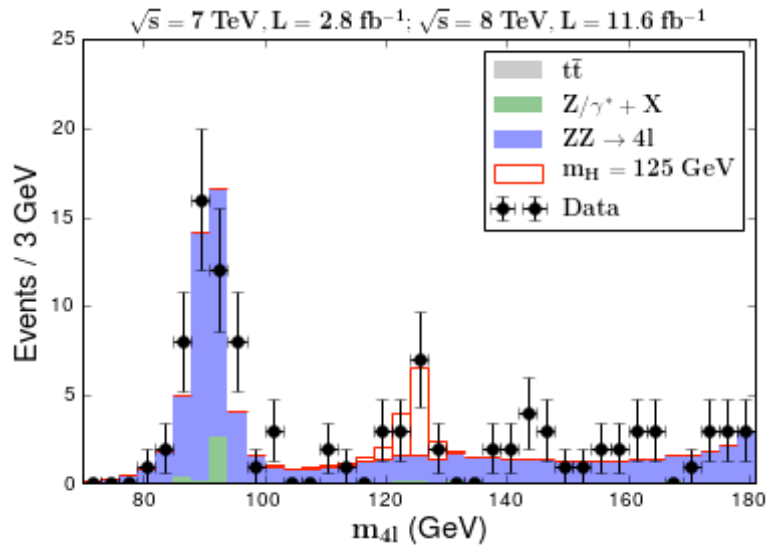

Figure 7: Actual CMS four-lepton result.

In 2018, nearly 300 four-lepton events became available, showing evidence of Zγ and ZZ events as well as a far smaller sample close to the Higgs mass. Work is underway to incorporate the new data so that students can make meaningful two- and four-lepton mass plots from their masterclass analysis.

## 3.2. Belle II

Belle II is an upgraded, more capable version of the Belle experiment at KEK to study B physics coming from electron-positron collisions. Based on the highly successful KEK B-Lab for students, the Belle II masterclass does not focus so much on individual events as on large numbers of events. The student task, then, is to use code similar to SCRATCH in order to filter data and make a variety of plots. The Belle II masterclass was piloted in International Masterclasses 2019, improved, and is expected to be available for 2020. [5]

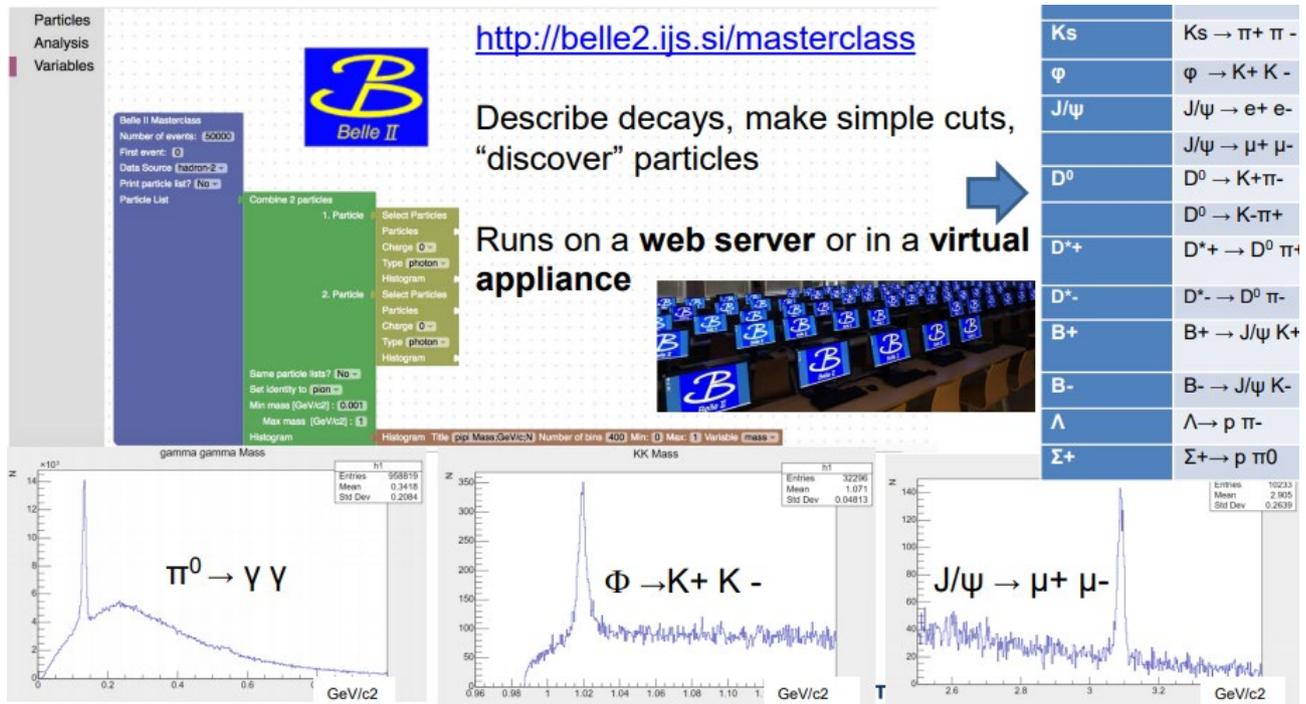

Figure 8: Belle II masterclass computing environment and results.

## 3.3. Particle Therapy

Medical physics has great potential to capture the interest of students who are more oriented toward life sciences or are more motivated by applications of science to society. The particle therapy masterclass is positioned to do just that. Student participants examine cancer scenarios and plan treatment with photon, carbon nucleus, and proton therapies, calculating optimal doses using the open source matRad toolkit. The particle therapy masterclass will be introduced into International Masterclasses in 2020. [6]

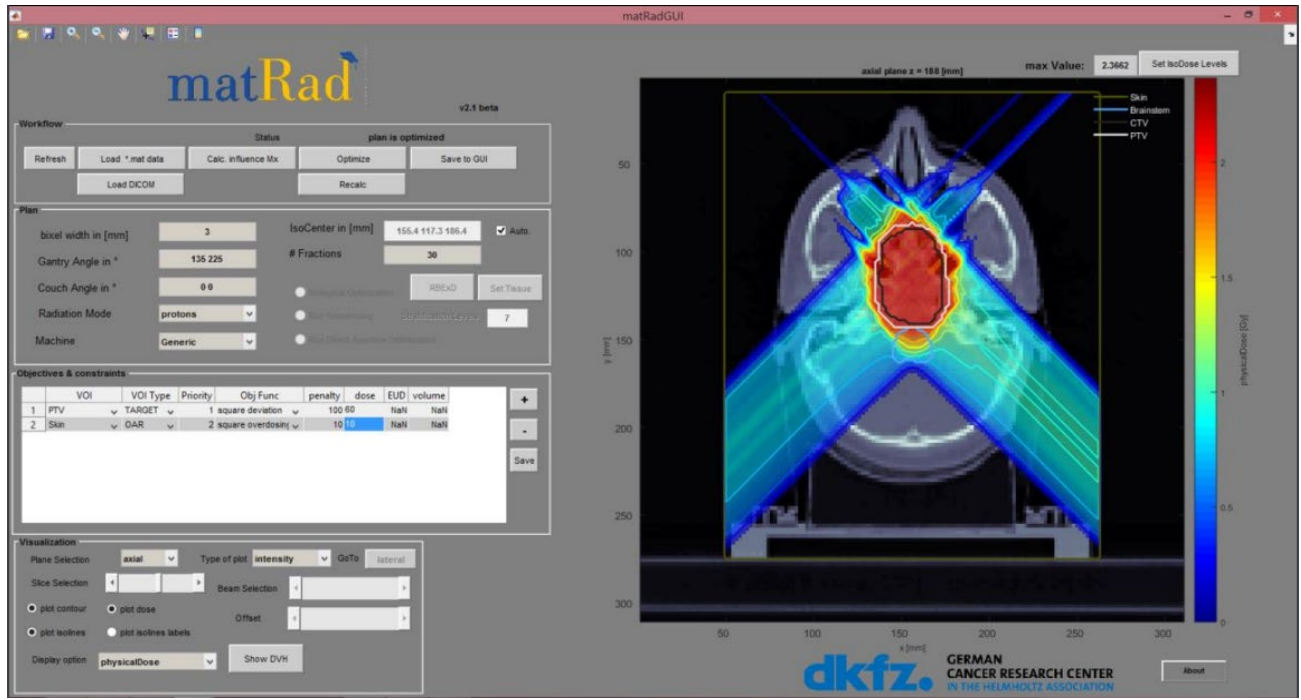

Figure 9: matRad particle therapy simulation.

### 3.4. Neutrino Masterclasses

As Fermilab and other institutions have begun to push the intensity frontier with accelerator-based neutrino experiments, physicists and educators became more interested in masterclasses to bring this physics to teachers and students. To this end, collaborators at the University of Rochester began to develop a measurement about five years ago to enable students to apply momentum studies to interactions of neutrinos with carbon nuclei in the Fermilab MINERvA experiment. [7] In 2018 and 2019, QuarkNet, Rochester physicists, and Fermilab developed this further into a full masterclass, which debuted and was tested in International Masterclasses 2019. Participants examine events in the MINERvA detector, first separating signal from background and then using conservation of momentum and the uncertainty principle to test models of the interaction. The more successful model enables students to find the neutrino beam energy, Fermi motion of neutrons in the nucleus, and an approximation of the carbon nuclear radius. Student success at the measurement and interest were both high. [8]

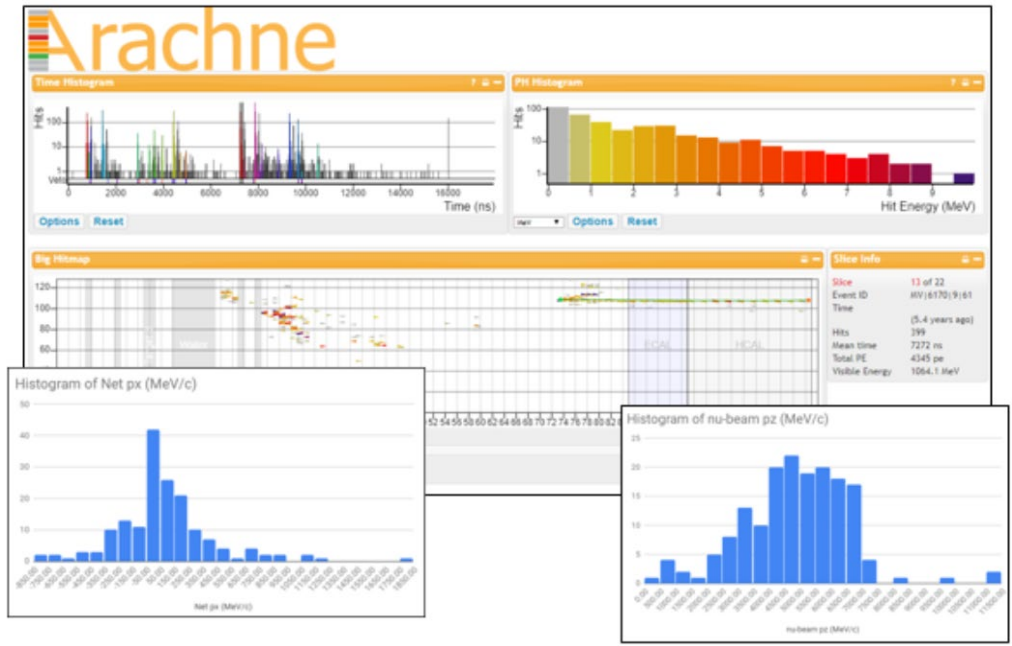

Figure 10: Arachne event display in MINERvA masterclass momentum histograms.

At the same time, Fermilab MicroBooNE physicists have been collaborating with QuarkNet to develop a detector commissioning masterclass. The liquid argon time-projection chamber (TPC) technology in MicroBooNE is sophisticated and sensitive; students can, for example, examine cosmic ray muon tracks in the detector to determine the purity of the argon and the overall response of the detector. The MicroBooNE masterclass will be tested in International Masterclasses 2020.

MINERvA tests models of neutrino interactions that physicists need to understand very well to make best use of DUNE when it comes online; MicroBooNE uses and enables student to test the key liquid argon TPC technology on which DUNE will rely.

This activity has spurred interest from other neutrino experiments so that several more are expected to come online in the coming years. DUNE physicists are already considering a simulated supernova measurement based on the projected capabilities of the detector.

## Conclusion: Bright, Interesting Future

CERN and Fermilab are currently the two main centers for moderating masterclass videoconferences. In addition, TRIUMF runs a single videoconference each year for mostly local British Columbia masterclass institutes. In 2019, Fermilab added to its portfolio with MINERvA masterclasses, and KEK moderated a single videoconference for the Belle II masterclass—the first from that laboratory. As the new masterclasses increase and expand, KEK and Fermilab will increase their activity while new videoconference centers will develop.

LHC masterclasses will continue to develop, as seen with the revision of the CMS masterclass and a planned new J/Ψ measurement for ALICE.

The new opportunities seen with W2D2 and IDWGS continue to be strong and new opportunities will develop.

## Acknowledgments

The authors wish to thank Uta Bilow, Technische Universität Dresden, for important material used in this paper. Work supported by the National Science Foundation and the International Particle Physics Outreach Group.